\documentclass[final,3p,times,twocolumn]{elsarticle}

\pdfoutput=1

\usepackage{amssymb}
\usepackage{amsmath}
\usepackage{graphicx}
\usepackage{dcolumn}
\usepackage{bm}
\usepackage{epsf}

\newcommand{\bra}[1]{\langle #1|}
\newcommand{\ket}[1]{|#1\rangle}

\newcommand{\la}{\left\langle}
\newcommand{\ra}{\right\rangle}

\begin{document}

\begin{frontmatter}



\title{Quantum work statistics of  linear and nonlinear  parametric oscillators}


\author{Sebastian Deffner \corref{cor}}
\ead{sebastian.deffner@physik.uni-augsburg.de}
\ead[URL]{www.physik.uni-augsburg.de/~deffnese}
\cortext[cor]{Corresponding author}

\author{Obinna Abah\corref{cor1}}

\author{Eric Lutz\corref{cor2}}

\address{Department of Physics, University of Augsburg, D-86135 Augsburg, Germany}

\begin{abstract}
We consider the nonequilibrium work distribution of a quantum oscillator with modulated angular frequency. We examine the discrete-to-continuous transition of the distribution as the temperature and the degree of nonadiabaticity of the frequency transformation are increased. We further develop a perturbative approach to analyze the effect of weak quartic anharmonicities, as well as of a random electric field on a charged oscillator. We find in both cases that  the degree of nonadiabaticity is enhanced by the perturbation.
\end{abstract}

\begin{keyword}
nanothermodynamics \sep nonequilibrium statistics \sep ion traps 

 \PACS 05.30.-d \sep 05.70.Ln \sep 05.40.-a


\end{keyword}

\end{frontmatter}


\section{Introduction\label{intro}}

The modern trend of miniaturization leads to the development of smaller and smaller devices, such as nanoengines  and molecular motors \cite{fen03,cer09,mar09}. On these very  short length scales, thermal as well as quantum fluctuations become important, and usual thermodynamic quantities, such as work and heat, acquire a stochastic nature. The traditional framework of thermodynamics, which  neglects fluctuations,  thus fails to provide a complete description of  nanosystems. Extensions of the second law to these small systems have been recently introduced in the form of the fluctuation theorem \cite{eva93,gal95} and the Jarzynski equality \cite{jar97}.   The Jarzynski  work relation,
\begin{equation}
\label{eq1}
\la \exp{\left(-\beta W\right)}\ra=\exp{\left(-\beta\Delta F\right)}
\end{equation}
permits determination of an equilibrium free energy difference $\Delta F$ from the fluctuations of the nonequilibrium work $W$ done on the system during an arbitrary transformation, quasistatic or not. The classical Jarzynski equality \eqref{eq1} is valid for both isolated and open systems \cite{jar02}. 
 It should be noted that the  system is initially  assumed to be   in a thermal state with inverse temperature $\beta$. However,  it is not required to be in an equilibrium state at the end of the transformation; for a general process, the system may  be arbitrarily far  from equilibrium.  For the case of an open system, the final free energy is  that of the asymptotic state reached by the system after equilibration with the heat bath, while for an isolated system, it corresponds to that of the hypothetical equilibrium state that would be obtained after coupling to the  bath. In Eq.~\eqref{eq1} the average of the exponentiated total work is taken over an ensemble of many realizations of the same process,  the nonequilibrium fluctuations of the  work being characterized by the probability distribution ${\cal P}(W)$. The classical Jarzynski equality has been verified experimentally using stretched biopolymers  \cite{lip02},  a mechanical torsion pendulum \cite{dou05}, and a colloidal particle in an anharmonic trap \cite{bli06}. At the quantum level, the Jarzynski relation has been shown to hold for isolated \cite{tas00,muk03} as well as for open systems \cite{tal08,cam09}, but an experimental investigation is still lacking. This is partly due to the fact that quantum work is not an observable in the usual sense, as it is not described by a Hermitian operator in Hilbert space, but by a two-time correlation function \cite{tal07}. A measurement scheme to determine the full quantum work statistics ${\cal P}(W)$, and hence verify the quantum Jarzynski equality, using ultracold trapped ions has been proposed in Ref.~\cite{hub08}. We have recently  explicitly  evaluated the continuous envelop of the work distribution ${\cal P}(W)$  for  a time-dependent harmonic oscillator, an analytically solvable model \cite{def08}. In the general quantum case, however, the work distribution is discrete, reflecting the quantized nature of the energy spectrum. In the present article, we provide a detailed discussion of the transition from discrete to  continuous distributions by introducing the cumulative work function. We further analyze the effect of a weak anharmonicity on ${\cal P}(W)$ by using time-dependent perturbation theory. We finally examine the effect of a weak random electric field on the work statistics of a charged harmonic oscillator, a case of high relevance for the thermodynamic study of realistic modulated Paul traps \cite{lei03}.

\section{Time-dependent harmonic oscillator \label{harm}}
We begin by reviewing the solution of the quantum parametric oscillator, following the method developed by Husimi  \cite{hus53}. We use this opportunity to extend the results of Ref.~\cite{hus53} and correct some misprints. The Hamiltonian of  a quantum mechanical harmonic oscillator with time-dependent angular frequency $\omega_t$ is
\begin{equation}
\label{eq2}
H_t=\frac{p^2}{2M}+\frac{M}{2}\omega^2_t x^2 \ .
\end{equation}
The parameterization $\omega_t$ starts at initial value $\omega_0$ at $t=t_0$ and ends at final value $\omega_1$ at $t=\tau$. We denote by $\phi_n^t$ the instantaneous eigenfunctions and by $E_n^t= \hbar\omega_t\left(n+1/2\right)$  the instantaneous eigenvalues  of the quadratic Hamiltonian (\ref{eq2}). The dynamics of the harmonic oscillator is Gaussian for any $\omega_t$. By introducing the  Gaussian wave function ansatz,
\begin{equation}
\label{eq3}
\psi_t (x)=\exp{\left(\frac{i}{2\hbar}\left[a_t x^2+2b_t x+c_t\right]\right)}, 
\end{equation}
the  Schr\"odinger equation for the  oscillator can be reduced to  a system of three coupled differential equations for the time-dependent coefficients $a_t$, $b_t$ and $c_t$,
\begin{eqnarray}
\label{eq4}
\frac{1}{M}\frac{da_t}{dt}&=&-\frac{1}{M^2}\,a^2_t-\omega^2_t,\\
\label{eq5}
\frac{db_t}{dt}&=&-\frac{1}{M}\,a_t b_t  ,\\ 
\label{eq6}
\frac{dc_t}{dt}&=&\frac{i\hbar}{M}\,a_t-\frac{1}{M}\,b^2_t  .
\end{eqnarray}
The nonlinear equation (\ref{eq4}) is of the Riccati type and is therefore solvable. It can be mapped to the equation of motion of a classical  time-dependent harmonic oscillator via $a_t=M\,\dot X_t/X_t$, and we obtain
\begin{equation}
\label{eq7}
\frac{d^2}{dt^2}\,X_t+\omega^2_t\,X_t=0 \ .
\end{equation}
With the solutions of (\ref{eq4})-(\ref{eq7}) the Gaussian wave function  $\psi_t(x)$ (\ref{eq3}) is fully characterized by the time-dependence of the angular frequency $\omega_t$. The general form of the propagator can be determined from $\psi_t (x)$ by noting that
\begin{equation}
\label{eq8}
\psi_t (x) = \int dx_0 \,U_{t,t_0}(x|x_0)\, \psi_{t_0}(x_0) \ .
\end{equation}
It is explicitly given by \cite{hus53}
\begin{equation}
\label{eq9}
U_{t,t_0} = \sqrt{\frac{M}{2\pi i \hbar X_t}} \exp{\left(\frac{i M}{2\hbar X_t} \left[\dot X_t x^2-2 xx_0+ Y_t x_0^2\right] \right)}\ ,
\end{equation}
where $X_t$ and $Y_t$ are solutions of Eq.~(\ref{eq7}) satisfying the boundary conditions $X_0=0$, $\dot X_0=1$ and $Y_0=1$, $\dot Y_0=0$, the latter being an expression of the commutation relation between position and momentum. 

\subsection{Method of generating functions}
The time variation of the  angular frequency (\ref{eq2}) induces transitions between different energy eigenstates of the oscillator. We are thus interested in the transition probabilities $p_{m,n}^\tau$ from an initial state $\ket{n}$ at $t_0=0$ to a final state $\ket{m}$ at $t=\tau$. In the following, we use the method of generating functions  to evaluate  $p_{m,n}^\tau$ \cite{hus53}. We start with the definition,
\begin{equation}
\label{eq10}
 p_{m,n}^{\tau}=\left|\int dx_0\int dx \, {\phi^*}_m^\tau(x)\, U_{\tau,0}(x|x_0)\,\phi^0_n(x_0)\right|^2\,,
\end{equation}
and denote the complex conjugate of a number $z$ by ${z}^*$. The quadratic generating function of $\phi_n^t(x)$ is
\begin{equation}
\label{eq11}
\begin{split}
\sum\limits_{n=0}^\infty\,u^n\,\phi_n^t(x)\,{\phi^*}^t_n(x_0)=\hspace{4cm} \\ \sqrt{\frac{M\omega_t}{\hbar\pi(1-u^2)}}\exp\left(-\frac{M\omega_t}{\hbar}\frac{(1+u^2)(x^2+x_0^2)-4u xx_0}{2(1-u^2)}\right),
\end{split}
\end{equation}
which can be calculated by a Fourier~expansion of the left-hand side of Eq.~(\ref{eq11}). The generating function $p_{m,n}^\tau$ is then defined as
\begin{equation}
\label{eq12}
P(u,v)=\sum\limits_{m,n}\,u^m v^n p_{m,n}^\tau \ .
\end{equation}
Combining Eqs.~(\ref{eq10}) and (\ref{eq11}), we find
\begin{equation}
\label{eq13}
\begin{split}
 P(u,v)=\hspace{4cm}\\
\frac{\sqrt2}{\sqrt{Q^{*}(1-u^2)(1-v^2)+(1+u^2)(1+v^2)-4uv}}\,.
\end{split}
\end{equation}
 The $(u,v)$-dependence of the generating function $P(u,v)$ remains the same for all possible transformations $\omega_t$. Details about the specific parameterization of the angular frequency only enter through different numerical values of the parameter $Q^*$ given by, 
\begin{equation}
\label{eq14}
Q^{*}=\frac{1}{2\omega_0\omega_1}\left\{\omega_0^2\,\left[\omega_1^2\,X^2_\tau+\dot X^2_\tau\right]+\left[\omega_1^2\,Y^2_\tau+\dot Y^2_\tau\right]\right\}. 
\end{equation}
From a physical point of view,  $Q^*$ can be regarded as a measure of the degree of adiabaticity of the process and will be discussed in more detail in the following subsection. Among the properties of the generating function $P(u,v)$, it is worth mentioning that the law of total probability $\sum\limits_n p_{m,n}^\tau=1$ is fulfilled and is equivalent to 
\begin{equation}
p(u,1)=\frac{1}{1-u}\,.
\end{equation}
 For a constant frequency, $\omega_t\equiv\omega_0$, we note that the solution of Eq.~(\ref{eq7}) is given by 
\begin{equation}X_t=\frac{1}{\omega_0}\, \sin{\left(\omega_0\, t\right)}\,,\hspace{1em}\text{and}\,\hspace{1em}  Y_t=\cos{\left(\omega_0\,t\right)}\, .
\end{equation}
The latter  imply with Eq.~(\ref{eq14}) that $Q^*= 1$ and  Eq.~(\ref{eq13}) thus simplifies to
\begin{equation}
P(u,v)\big|_{Q^*=1}=\frac{1}{1-uv}\,,
\end{equation}
which is equivalent to $p_{m,n}^\tau=\delta_{m,n}$, indicating the absence of  transitions, as expected. The symmetry relation of the generating function (\ref{eq13}), $P(-u,-v)=P(u,v)$, further shows that $p_{m,n}^\tau=0$ if $m$, $n$ are of different parity. This is an expression of a selection~rule $m=n\pm2k$, where $k$ is an integer. In subsection \ref{ideal}, we rederive this selection rule by means of time-dependent perturbation theory. We mention in addition that the transition probabilities are symmetric, $p_{m,n}^\tau=p_{n,m}^\tau$, following $P(u,v)=P(v,u)$. Explicit expressions for the transitions probabilities $p_{m,n}^\tau$ are given in terms of hypergeometric functions in \ref{trans}.

\subsection{Measure of adiabaticity \label{measure}}
The parameter $Q^*$ defined in (\ref{eq14}) can be given a simple physical meaning \cite{hus53}. We base our discussion of adiabaticity on the equivalent classical harmonic oscillator (\ref{eq7}) since the generating function $P(u,v)$ (\ref{eq13}) is fully determined through the classical  solutions $X_t$ and $Y_t$. For an adiabatic transformation, the action of the oscillator, given by the ratio of the energy to the angular frequency, is a time-independent constant. For quasistatic processes we have the two adiabatic invariants,
\begin{equation}
\label{eq15}
\frac{\dot X^2_t + \omega^2_t X^2_t}{\omega_t}=\frac{1}{\omega_0},\hspace{1em}\text{and}\hspace{1em} \frac{\dot Y^2_t + \omega^2_t Y^2_t}{\omega_t}=\omega_0. 
\end{equation}
From the definition (\ref{eq14}) of the parameter $Q^*$, we see that in this case we simply have $Q^*=1$. As mentioned earlier this implies $P(u,v)=(1-u v)^{-1}$ and $p^\tau_{m,n}= \delta_{m,n}$. The latter  is an expression of the quantum adiabatic theorem: For  infinitely slow transformations no transitions between different quantum states occur. For fast transformations, on the other hand,  we can regard $Q^*$ as a measure of the degree of nonadiabaticity of the process. As an illustration, we  evaluate mean and variance of the energy of the oscillator at time $\tau$ and express them as a function of $Q^*$. For a transition from initial state $\ket{n}$ to final state $\ket{m}$, the mean-quantum number of the final state $\la m\ra_n$ can be obtained by taking  the first derivative of the generating function (\ref{eq13}) of  $p_{m,n}^\tau$,
\begin{equation}
\label{eq16}
\begin{split}
 \sum\limits_n u^n\sum\limits_m m \,p_{m,n}^\tau=&\frac{\partial P(u,v)}{\partial
v}\bigg|_{v=1}\\
=&\frac{Q^*(1+u)-(1-u)}{2(1-u)^2},
\end{split}
\end{equation}
and expanding the left hand side of (\ref{eq16})   in powers of $u$:
\begin{equation}
\label{eq17}
\la m\ra_n = \sum_m m \,p_{m,n}^\tau = \left(n+\frac{1}{2}\right) Q^* -\frac{1}{2}.
\end{equation}
Noting that   $p_n^0=\exp(-\beta E_n^0)/Z_0$, the mean energy of the oscillator at time $\tau$ then reads
\begin{equation}
\label{eq18}
\begin{split}
 \la H_\tau\ra=\sum\limits_{n}\hbar\omega_1\left(\la m\ra_n+\frac{1}{2}\right)\,p_n^0\\
=\frac{\hbar\omega_1}{2}\,Q^*\,\coth{\left(\frac{\beta}{2}\hbar\omega_0\right)}\,.
\end{split}
\end{equation}
Since $\la m\ra_n\geq0$, and hence $\la H_\tau\ra\geq\hbar\omega_1/2$, the parameter $Q^*$ necessarily satisfies   $Q^*\geq1$ for generic processes. In the zero temperature limit,
Eq.~(\ref{eq18}) reduces to
\begin{equation}
\la H_\tau\ra=\frac{\hbar\omega_1}{2}\,Q^* 
\end{equation}
The above equation  corrects a misprint appearing  in Eq.~(5.21) of Ref.~\cite{hus53} ($\omega_0$ should be replaced by   $\omega_1$). The mean-square quantum number $\la m^2\ra_n-\la
m\ra_n^2$ at time $\tau$ can be calculated in a similar way by considering $\la m(m-1)\ra_n$. By differentiating Eq.~(\ref{eq13}) twice, we have 
\begin{eqnarray}
\label{eq19}
 \sum\limits_n u^n \sum\limits_m m(m-1)\,p_{m,n}^\tau = \hspace{2cm}\nonumber\\
\frac{1+(u-6)u+3 {Q^*}^2(u+1)^2+4 Q^*(u^2-1)}{4(u-1)^3}\ ,
\end{eqnarray}
and  a series expansion  in powers of $u$ leads to
\begin{equation}
\label{eq20}
\begin{split}
 \la~m(m-~1)~\ra_n=&\frac{1}{4}\,\Big[1-2n\,\left(n+1\right)-4 \,Q^*\left(2n
+1\right)\\
+& 3 \,{Q^*}^2\left(2n^2 + 2n+ 1\right)\Big].
\end{split}
\end{equation}
The mean-square quantum number is then obtained by combining Eqs.~(\ref{eq17}) and
(\ref{eq20}):
\begin{equation}
\label{eq21}
 \la m^2\ra_n-\la m\ra_n^2 = \frac{1}{2} \left({Q^*}^2 -1\right)\left(n^2+n+1\right).
\end{equation}
From Eqs.~(\ref{eq17}), (\ref{eq18}) and
(\ref{eq21}), we can finally write  the variance of the energy as
 \begin{equation}
\label{eq22}
\begin{split}
 \la H^2_\tau\ra-{\la
H_\tau\ra}^2=\frac{\hbar^2\omega_1^2}{8}\,\text{csch}^2{\left(\frac{\beta}{2}\,\hbar\omega_0\right)}\hspace{1cm}\\
\times\,\left(1-4 {Q^*}^2-3
\cosh{\left(\beta\hbar\omega_0\right)}+4 Q^*\sinh{\left(\beta\hbar\omega_0\right)}\right)\,.
\end{split}
\end{equation}
The  zero-temperature limit,
\begin{equation}
 \la H^2_\tau\ra-{\la
H_\tau\ra}^2=\frac{\hbar^2\omega_1^2}{2}\,\left({Q^*}^2-1\right)\ ,
\end{equation}
is again the correct version of Eq.~(5.22) of Ref.~\cite{hus53}. Equation (\ref{eq21}) indicates that   the parameter $Q^*$
directly controls the magnitude of the variance of the occupation number,  $ \la
m^2\ra_n-\la m\ra_n^2$. In the adiabatic limit, where $Q^*=1$, we readily  get $\la
m\ra_n=n $ and $\la m^2\ra_n-\la m\ra_n^2=~0$. We therefore recover that for
adiabatic  processes the  system remains in its initial state, $\ket{m}=\ket{n}$. On
the other hand, for fast nonadiabatic processes, the mean $\la m\ra_n$ and the
dispersion $\la m^2\ra_n-\la m\ra_n^2 $ increase with increasing values of $Q^*$,
indicating that the quantum oscillator ends in a final state $\ket{m}$  which is
farther and farther away from the initial state $\ket{n}$. The latter corresponds to  larger and larger final values of the  mean energy and energy variance, Eqs.~\eqref{eq18} and \eqref{eq22}. 

It is worth mentioning that the above discussion of the  adiabaticity parameter $Q^*$  for the parametric oscillator is close in spirit to  the Einstein criteria for adiabatic processes \cite{kul57}. Einstein noted that 
 for an adiabatic process, the classical action of the oscillator, $\la H_t\ra/\omega_t$, should remain
constant and the number of quanta should therefore  remain unchanged. In the present situation, we have $\la H_\tau\ra/\omega_\tau\propto Q^*$, and the action only remains constant when $Q^*=1$. The latter is precisely the condition that we derived for an adiabatic transformation. For nonadiabatic processes, the parameter
$Q^*>1$ thus gives a measure for the increase of the classical action of the oscillator.  Further discussions of adiabatic measures can be found in Ref.~\cite{tak92}.

\section{Work probability density function \label{prob}}
In this section, we introduce the probability distribution ${\cal P}(W)$ on the nonequilibrium work done on the parametric oscillator during a variation of its angular frequency.  We give the expressions of its continuous envelop in the limit of high and low temperatures, for adiabatic and nonadiabatic transformations \cite{def08}. We further compare our nonadiabatic results with those recently derived by van Zon and coworkers in Ref.~\cite{zon08}.  

In quantum mechanics, the probability density function of  work is obtained by considering the  difference between final and initial  energy eigenstates, $E_{m}^{\tau }$ and $E_{n}^{0}$, averaged over all possible final and initial states. The probability distribution  of the total work done  during time $\tau$ can thus be written as \cite{tal07},
\begin{equation}
\label{eq23}
{\cal P}(W)=\sum\limits_{m,n}\,\delta\left(W-(E_{m}^{\tau }-E_{n}^{0})\right)\,p_{m,n}^{\tau}\, p^0_{n},
\end{equation}
where $p_n^0=\exp(-\beta E_n^0)/Z_0$ is the initial (thermal) occupation probability. Equation \eqref{eq23} show that work is  a random quantity because of the presence of  both thermal and quantum uncertainties, encoded respectively in $p_n^0$ and $p_{m,n}^{\tau}$. The characteristic function of the work, defined as the Fourier transform of the  probability distribution (\ref{eq23}), 
\begin{equation}
\label{eq24}
 G(\mu) = \int dW \exp{\left(i \mu W\right)} \,{\cal P}(W)\,,
\end{equation}
can be written in closed form in terms of the energies, $\varepsilon_0=\hbar\omega_0$, $\varepsilon_1=\hbar\omega_1$ ($\Delta\varepsilon=\varepsilon_1-\varepsilon_0$). Using  expression (\ref{eq13}) of the generating function, we have
\begin{equation}
\label{eq25}
 G(\mu)=\sqrt{\frac{2}{n\left(Q^*\right)}}\left(1-\exp{\left(-\beta\varepsilon_0\right)}\right)\,\exp{\left(i\mu \,\frac{\Delta\varepsilon}{2}\right)}
\end{equation}
where the denominator is given by
\begin{equation}
\label{eq26}
\begin{split}
n\left(Q^*\right)=& Q^{*}\left(1-\exp{\left(2i \mu \varepsilon_1\right)}\right)\left(1-\exp{\left(-2(i \mu +\beta)\varepsilon_0\right)}\right)\\
+&\left(1+\exp{\left(2i \mu \varepsilon_1\right)}\right)\left(1+\exp{\left(-2(i \mu +\beta)\varepsilon_0\right)}\right)\\
-&4\,\exp{\left(i \mu \varepsilon_1\right)} \exp{\left(-(i \mu +\beta)\varepsilon_0\right)}.
\end{split}
\end{equation}
The above equations for $G(\mu)$ are exact and fully characterizes the work distribution of the time-dependent quantum harmonic oscillator (\ref{eq2}) for arbitrary parameterizations of the angular frequency $\omega_t$. As mentioned previously, different realizations of $\omega_t$ merely lead to different values of the parameter $Q^*$. The direct analytic evaluation of the nonequilibrium work distribution ${\cal P}(W)$ by Fourier inverting Eq.~(\ref{eq25}) is not feasible in the general case due to the nonanalytic denominator of $G(\mu)$. We here provide analytical approximations in various limits of interest. The detailed derivation of the following probability distributions  can be found in Ref.~\cite{def08}. 

\paragraph{Adiabatic transformations}
As discussed  in subsection \ref{measure},   adiabatic transformations are characterized by  $Q^*=1$. By approximating Eq.~\eqref{eq25} in the limit of zero-temperature, $\hbar\beta\gg 1$, we get
\begin{equation}
\label{eq27}
{\cal P}(W)=\delta\left(W-\frac{\Delta\varepsilon}{2}\right)\, . 
\end{equation}
Equation (\ref{eq27}) expresses the deterministic nature of adiabatic processes at zero-temperature: the oscillator  starts and ends in its ground state.  Work is hence simply given by the difference of final and initial ground state energies. 
In the opposite classical limit, $\hbar\beta\ll 1$, we find
\begin{equation}
\label{eq28}
{\cal P}(W)=\frac{\beta\omega_0}{\Delta\omega}\exp{\left(-\frac{\beta\omega_0}{\Delta\omega}\,W\right)}\,\Theta(W).
\end{equation}
Equation~(\ref{eq28}) is identical to the classical work probability distribution derived by Jarzynski \cite{jar97a}. Note that in this case only positive work fluctuations occur, whose magnitude is controlled by the finite temperature.
\paragraph{Nonadiabatic transformations}
We next  consider nonadiabatic transformation which correspond to $Q^* >1$. In the zero-temperature limit, $\hbar\beta\gg 1$, the work distribution can be approximated in the limit of small $\varepsilon_1$ by, 
\begin{equation}
\label{eq29}
{\cal P}(W)=\frac{\exp{\left(-\frac{W-\Delta \varepsilon/2}{(Q^*-1)\varepsilon_1}\right)}}{\sqrt{\pi (Q^*-1) \,\varepsilon_1\,(W-\Delta \varepsilon/2)}}\ .
\end{equation}
The zero-temperature nonadiabatic work distribution (\ref{eq29}) is valid when $W\geq\Delta\varepsilon/2$. This condition follows from the existence  of the minimal ground state energy of the oscillator. An expansion of Eq.~(\ref{eq25})  in the classical limit, $\hbar\beta\ll 1$, leads to
\begin{equation}
\label{eq30}
\begin{split}
 {\cal P}(W)=&\sqrt{\frac{\beta^2\omega_0^2/\pi^2}{2 Q^*\omega_0\omega_1-\omega_0^2-\omega_1^2}}\\
\times&\exp{\left(\frac{Q^*\omega_0\omega_1-\omega_0^2}{2 Q^*\omega_0\omega_1-\omega_0^2-\omega_1^2}\,\beta W\right)}\\
\times& K_0\left(\frac{\beta\omega_0\omega_1\,\sqrt{{Q^*}^2-1}}{\left|2Q^*\omega_0\omega_1-\omega_0^2-\omega_1^2\right|}\,|W|\right),
\end{split}
\end{equation}
where $\Gamma(x)$ denotes the Euler Gamma function and $K_{\nu}(x)$ is the Macdonald function, that is the modified Bessel function of the third kind. Equation \eqref{eq30} is well-defined provided the term under the square-root is positive. This implies that the parameter $Q^*$ is larger than the value $Q^*_{\text{ss}}$ that corresponds to a sudden (instantaneous) switch of the frequency from $\omega_0$ to $\omega_1$,
\begin{equation}
\label{eq31}
Q^*>\frac{\omega_0^2+\omega_1^2}{2 \omega_0\omega_1}=Q^*_{\text{ss}}\,.
\end{equation}
 An example for a frequency parameterization resulting in an unbounded, divergent value of $Q^*$ can be found in Ref.~\cite{gal09}. In the regime where $Q^*<Q^*_{\text{ss}}$, the work distribution has been derived by   van Zon and coworkers  by explicitly evaluating the complex integral of the inverse Fourier transform \cite{zon08}. It reads in our notation
\begin{equation}
\label{eq32}
\begin{split}
 {\cal P}(W)=&\sqrt{\frac{\beta^2\omega_0^2}{\omega_0^2-2 Q^*\omega_0\omega_1+\omega_1^2}}\\
\times&\exp{\left(\frac{Q^*\omega_0\omega_1-\omega_0^2}{2 Q^*\omega_0\omega_1-\omega_0^2-\omega_1^2}\,\beta W\right)}\\
\times& I_0\left(\frac{\beta\omega_0\omega_1\,\sqrt{{Q^*}^2-1}}{\left|2Q^*\omega_0\omega_1-\omega_0^2-\omega_1^2\right|}\,|W|\right)\Theta(W)\,,
\end{split}
\end{equation}
where $I_0 (x)$ denotes the modified Bessel function of the first kind. It is worth mentioning that, in contrast to the adiabatic  \eqref{eq28} and nonadiabatic \eqref{eq32} results,   negative work values can occur in Eq.~(\ref{eq30})   for large values of  the parameter $Q^*$ \cite{zon08}.

\begin{figure}
\centering
\includegraphics[width=0.47\textwidth]{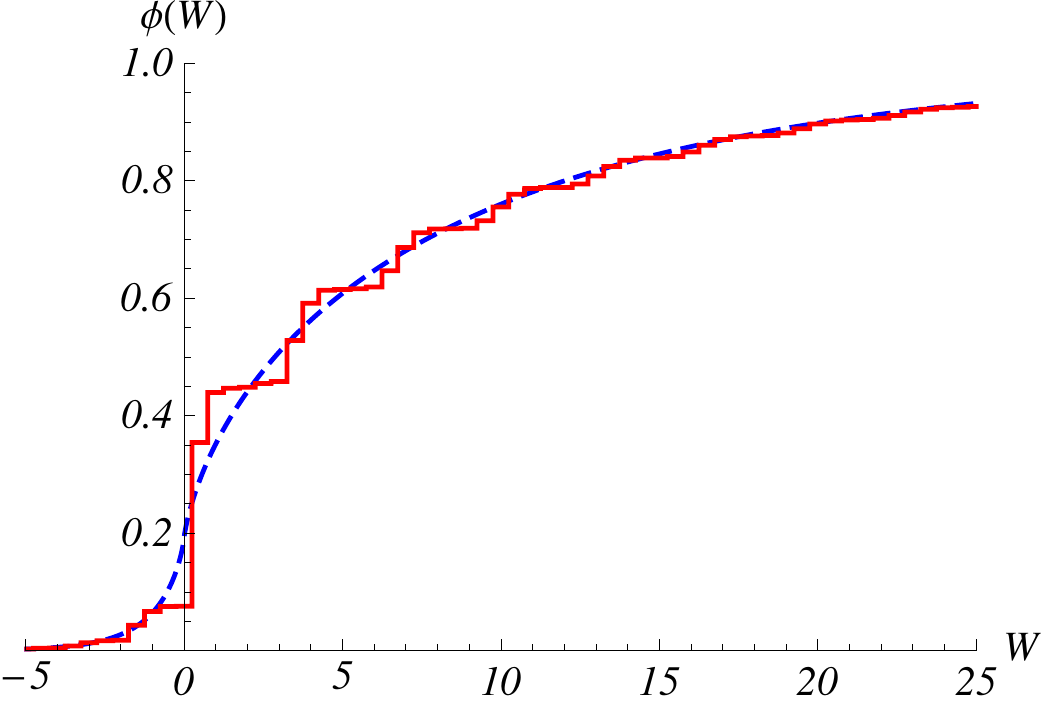}
\caption{Cumulative work distribution  $\phi(W)$ (\ref{eq33}) for  the exact quantum expression (\ref{eq23}) (red) and the approximated  classical form (\ref{eq30}) (blue, dashed) for the parameters $\omega_0=1$, $\omega_1=1.5$, $\hbar=1$, $\beta=0.5$ and $Q^*=3$. The discrete structure of the quantum distribution is easily recognized as compared to the continuous classical distribution.\label{fig1}}
\end{figure}

\section{Transition between discrete and continuous nonequilibrium work distributions\label{discr}}

We now  turn to the analysis of the transition from the discrete work probability density \eqref{eq23} to the continuous   analytical approximations Eqs.~(\ref{eq28}), (\ref{eq29}), (\ref{eq30}) and (\ref{eq32}), and provide a criterion for the observation of the 
discreteness of ${\cal P}(W)$.
The definition (\ref{eq23}) makes clear that the nonequilibrium work distribution consists of a discrete sum of delta peaks when the energy spectrum of the system is quantized. This is in stark contrast to classical work distributions which are continuous. The discreteness of  ${\cal P}(W)$ can therefore be regarded as the hallmark of the quantum nature of work. In order to quantitatively investigate the discrete-to-continuous transition, we introduce the cumulative probability distribution $\phi(W)$ defined as,
\begin{equation}
\label{eq33}
 \phi(W)=\int\limits_{\chi}^{W}dW'\,{\cal P}(W') \ .
\end{equation}
In the above equation, the constant $\chi$ has to be chosen according to the range validity of the work distribution, e.g. in the zero-temperature limit $\chi=\Delta\varepsilon/2$. The cumulative distribution \eqref{eq33} is a staircase function in  the case of a discrete distribution ${\cal P}(W)$ and turns over to a smooth function in the continuous limit.  Figure  \ref{fig1} shows the exact quantum function $\phi(W)$ for a given set of parameters, together with the  continuous classical approximation corresponding to Eq.~(\ref{eq30}). The numerical value of the  parameter  $Q^*$ has been chosen  to describe realistic experiments with  ion traps \cite{hub08a}. In order to obtain a criterion for  the discreteness of the work distribution ${\cal P}(W)$, we consider   the mean  energy of the  harmonic oscillator as given by Eq.~(\ref{eq18}). The energy spectrum of the harmonic oscillator is usually considered to be quasi-continuous when the mean quantum number is much larger than the level separation \cite{mes61}, or in  other words, when the mean energy is much larger than the energy quantum, $\la H_\tau\ra \gg\hbar\omega_1$. An inspection of Eq.~(\ref{eq18}) reveals that this limit can be achieved in two ways:
\begin{figure}
\centering
\includegraphics[width=0.47\textwidth]{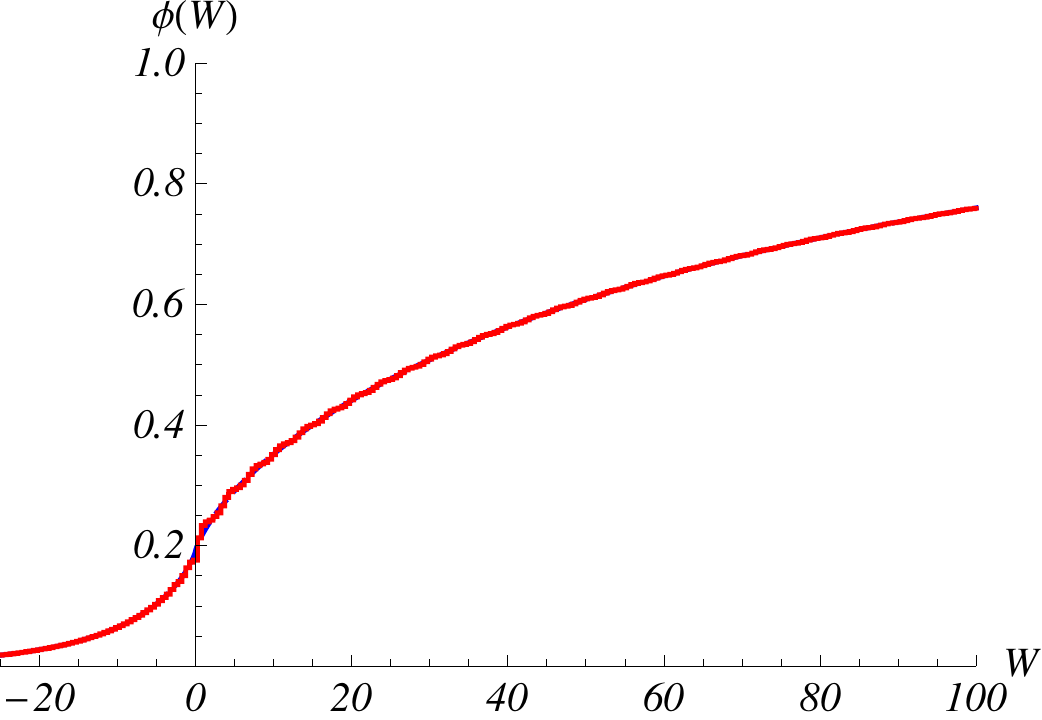}
\caption{In the limit of high temperatures,  the cumulative work function  $\phi(W)$ for the exact discrete case  (\ref{eq23})  (red) is indistinguishable from the continuous approximation (\ref{eq30}) (blue, dashed). Here $\omega_0=1$, $\omega_1=1.5$, $\hbar=1$, $\beta=0.05$ and $Q^*=3$.\label{fig2}} 
\end{figure}
\paragraph{High temperatures}
In the limit of high temperatures, $\hbar\beta\ll1$, we have $\coth{\left(\beta/2\,\hbar\omega_0\right)}\gg1$. For a fixed value of  $Q^*$, we therefore expect the nonequilibrium work distribution to become continuous when the temperature is increased. Figure \ref{fig2} depicts the  exact cumulative distribution $\phi(W)$ and  the  approximate result corresponding to Eq.~(\ref{eq30}) for the same parameters as in Fig.~\ref{fig1} except the temperature which has been increased by a factor ten; the two curves are indistinguishable.\paragraph{Large $Q^*$ values}
Another way to reach a large mean energy (\ref{eq18}) is to increase the value of $Q^*$, while keeping the temperature constant. The work distribution can thus become continuous when the degree of nonadiabaticity of the process is increased. 
Figure \ref{fig3} shows the exact and approximate cumulative distribution $\phi(W)$ for a  value of $Q^*$ twenty times larger  than in Fig.~\ref{fig1}, all other parameters being the same: the two curves are again not distinguishable.  It is interesting to notice that the transition to a continuous work distribution occurs faster when augmenting the temperature than  the degree of nonadiabaticity $Q^*$ (a factor two for the parameters of Fig.~\ref{fig1}), the reason being that the mean  energy (\ref{eq18})  does not depend linearly  on temperature in contrast to $Q^*$ (the temperature dependence becomes linear only for high temperatures).
\begin{figure}
\centering
\includegraphics[width=0.47\textwidth]{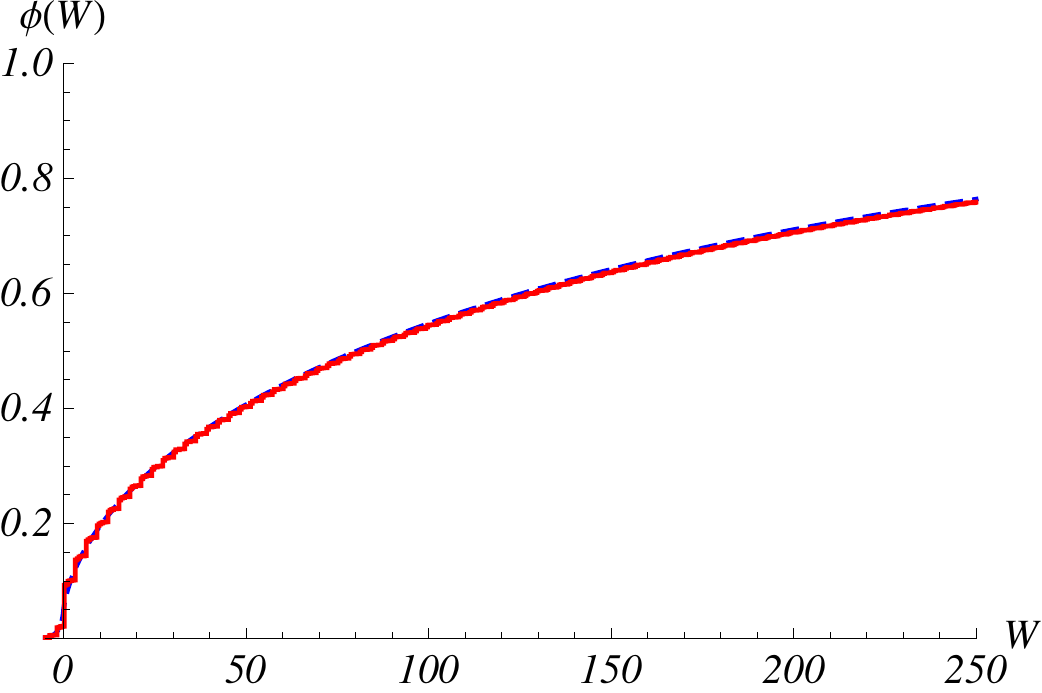}
\caption{In the limit of high $Q^*$ values,  the cumulative work function  $\phi(W)$ for the exact discrete case  (\ref{eq23})  (red) is indistinguishable from the continuous approximation (\ref{eq30}) (blue, dashed). Here $\omega_0=1$, $\omega_1=1.5$, $\hbar=1$, $\beta=0.5$ and $Q^*=60$.\label{fig3}} 
\end{figure}
\paragraph{Noninteger ratios $\varepsilon_1/\varepsilon_0$} 
One may wonder whether it is possible to have continuous work distributions, for any values of $Q^*$ and $\beta$, by for instance considering non-rational quotients $\varepsilon_1/\varepsilon_0$. In this limit, the gaps between $\delta$-peaks could be filled  and ${\cal P}(W)$ would appear continuous. This is not the case, however.  We note indeed that for any ratio $\varepsilon_1/\varepsilon_0$, we do not obtain real continuous distributions with respect to the real set. The $\delta$-functions in Eq.~(\ref{eq23}) only contribute to the total sum if their argument is zero. The only permitted work values are therefore of the form, 
\begin{equation}
\label{eq35}
W=\varepsilon_1\left(m+\frac{1}{2}\right)-\varepsilon_0\left(n+\frac{1}{2}\right)\,.
\end{equation}
Now since $m$ and $n$ are integers, only countably infinite values of $W$ can occur, implying that  the work distribution is discrete in any case. The latter follows from the fact that  rational numbers are densely distributed  within the real set.
\paragraph{Effect of the selection rule}
Finally, we examine the  effect of the  selection rule,
$m=n\pm2k$, noted in  section \ref{harm} on the work distribution ${\cal P}(W)$. The existence of this selection rule limits the possible values that the step sizes of the cumulative distribution $\phi(W)$ can take. For simplicity, we consider a harmonic oscillator which is initially prepared
in a given energy eigenstate $\ket{n_0}$. A direct consequence of Eq.~\eqref{eq35} and of the selection rule is then that the allowed work values are, 
\begin{equation}
W=\hbar\left(\omega_1-\omega_0\right)\,\left(n_0+\frac{1}{2}\right)\,\pm\,\hbar\omega_1
\, 2k\,,
\end{equation}
where $k$ are integers. For two neighboring work values, the step size is constant and reads,
\begin{equation}
\Delta W=2 \hbar\omega_1\,.
\end{equation}
In Fig.~\ref{fig4} we illustrate the effect of the selection rule by plotting the
corresponding cumulative distribution  with the same parameters as in
Fig.~\ref{fig1}. 

\begin{figure}
\centering
\includegraphics[width=0.47\textwidth]{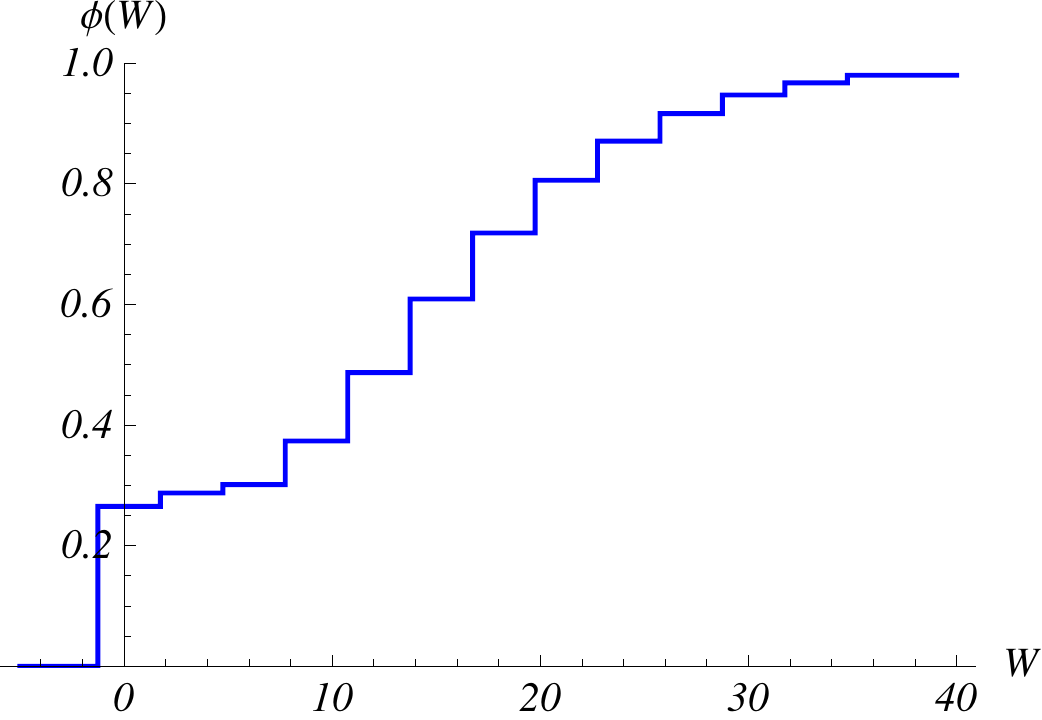}
\caption{Cumulative work distribution (\ref{eq33}) for an initially pure state $\ket{n_0}$, illustrating  the sum rule, $m=n\pm 2k$, which limits the possible sizes of the work steps ($n_0=3$, $\varepsilon_0=1$, $\varepsilon_1=1.5$
and $Q^*=3$).\label{fig4}}
\end{figure}

\section{Perturbation theory \label{pert}}
We develop in the present section a perturbative approach to determine the quantum work distribution ${\cal P}(W)$. We treat in detail the case of a small anharmonic correction to the potential, as well as the effect of a small external fluctuating electric field on a charged harmonic oscillator. Both situations are motivated by the experimental study of the quantum work statistics in linear Paul traps \cite{hub08}. We begin by giving a simple perturbative derivation of the sum rule of section  \ref{harm}, which does not require the full solution of the Schr\"odinger equation. 

\subsection{Selection rule \label{ideal}}

We start by  rewriting Hamiltonian (\ref{eq2})  as
\begin{equation}
\label{eq36}
H_t=H_0+\Omega_t \ ,
\end{equation}
where we have introduced the unperturbed Hamiltonian
\begin{equation}
\label{eq37}
H_0=\frac{p^2}{2M} +\frac{M}{2} \omega_0^2 x^2, 
\end{equation}
and a "perturbation" term\begin{equation}
\label{eq38}
\Omega_t=-\frac{M}{2}\, \left(\omega_0^2-\omega_t^2\right)\,x^2\,.
\end{equation}
The latter can be considered small for small frequency changes.
In first order time-dependent perturbation theory,  the transition probabilities between initial state $\ket{n}$ and  final state $\ket{m}$ are given by \cite{coh1},
\begin{equation}
\label{eq39}
 p_{m,n}^\tau=\left|\delta_{m,n}+\frac{1}{i\hbar}\int\limits_0^\tau dt\, \exp{\left(i\omega_{m,n} t\right)}\,\Omega_{m,n}^t\right|^2 \ ,
\end{equation}
where $\hbar\omega_{m,n}=E_m^0-E_n^0$ denotes the difference of the unperturbed energy eigenvalues and $\Omega_{m,n}^t=\bra{m}\Omega_t\ket{n}$ are the  corresponding interaction matrix elements. By expressing  the position operator, $x=\sqrt{\hbar/2 M \omega_0}\left(a^\dagger+a\right)$, in terms of the usual ladder operators, $a^\dagger\ket{n}=\sqrt{n+1}\,\ket{n+1}$ and $a\,\ket{n}=\sqrt{n}\,\ket{n-1}$,  the interaction matrix elements can be written explicitly as
\begin{equation}
\label{eq40}
\begin{split}
 \Omega_{m,n}^t=&-\frac{\hbar}{4\omega_0}\,\left(\omega_0^2-\omega^2_t\right)\,\Big[\sqrt{n+1}\sqrt{n+2}\,\delta_{m,n+2}\\
+&\,\left(2n+1\right)\,\delta_{m,n}\,+\,\sqrt{n-1}\sqrt{n}\,\delta_{m,n-2}\Big],
\end{split}
\end{equation}
with the Kronecker-delta $\delta_{m,n}$. Equation (\ref{eq40}) shows that only transitions that satisfy $m=n\pm2$ are possible, which is precisely the selection rule noted earlier. It should be emphasized that this selection rule is at variance with usual textbook examples which contain the selection rule $m=n\pm1$. The latter applies to a quantum oscillator driven by a small perturbation linear in the position, whereas we here deal with a perturbation (\ref{eq38}) which is quadratic in $x$. The full expression of the transitions probabilities (\ref{eq39}) that follow from Eq.~(\ref{eq40}) is given in \ref{trans_per}.

\subsection{Anharmonic corrections\label{anharmonic}}
A method to experimentally measure the quantum work distribution in modulated ion trap systems has been  proposed in Ref.~\cite{hub08}. In these systems, the confining potential is harmonic to a very good accuracy \cite{hub08a}. One attractive feature of linear Paul traps is however  the possibility to modify the shape of the potential with the help of external gate voltages. We here investigate the influence of a small quartic anharmonicity on the work distribution ${\cal P}(W)$.  As before, we write  the total Hamiltonian as
\begin{equation}
\label{eq41}
H_t = H_0 + \Omega_t + A_t\ ,
\end{equation}
where the first anharmonic correction is given by,
\begin{equation}
\label{eq42}
A_t=\alpha_t\, x^4\,.
\end{equation}
The total transition probabilities can then be written as 
\begin{equation}
\label{eq43}
p_{m,n}^{\tau} = \left| \delta_{m,n} + \frac{1}{i\hbar} \int\limits_{0}^{\tau} \, dt \, \exp{\left(i \omega_{m,n} \, t\right)}\, \left( \Omega_{m,n}^t + A_{m,n}^t \right) \right|^2\,,
\end{equation}
where  $A_{m,n}^t$ are the anharmonic interaction matrix elements. The analytic transition probabilities $p_{m,n}^{\tau}$ are again given in \ref{trans_per}. For the sake of clarity, we will continue with a numerical discussion of the results.
\begin{figure}
\centering
\includegraphics[width=0.47\textwidth]{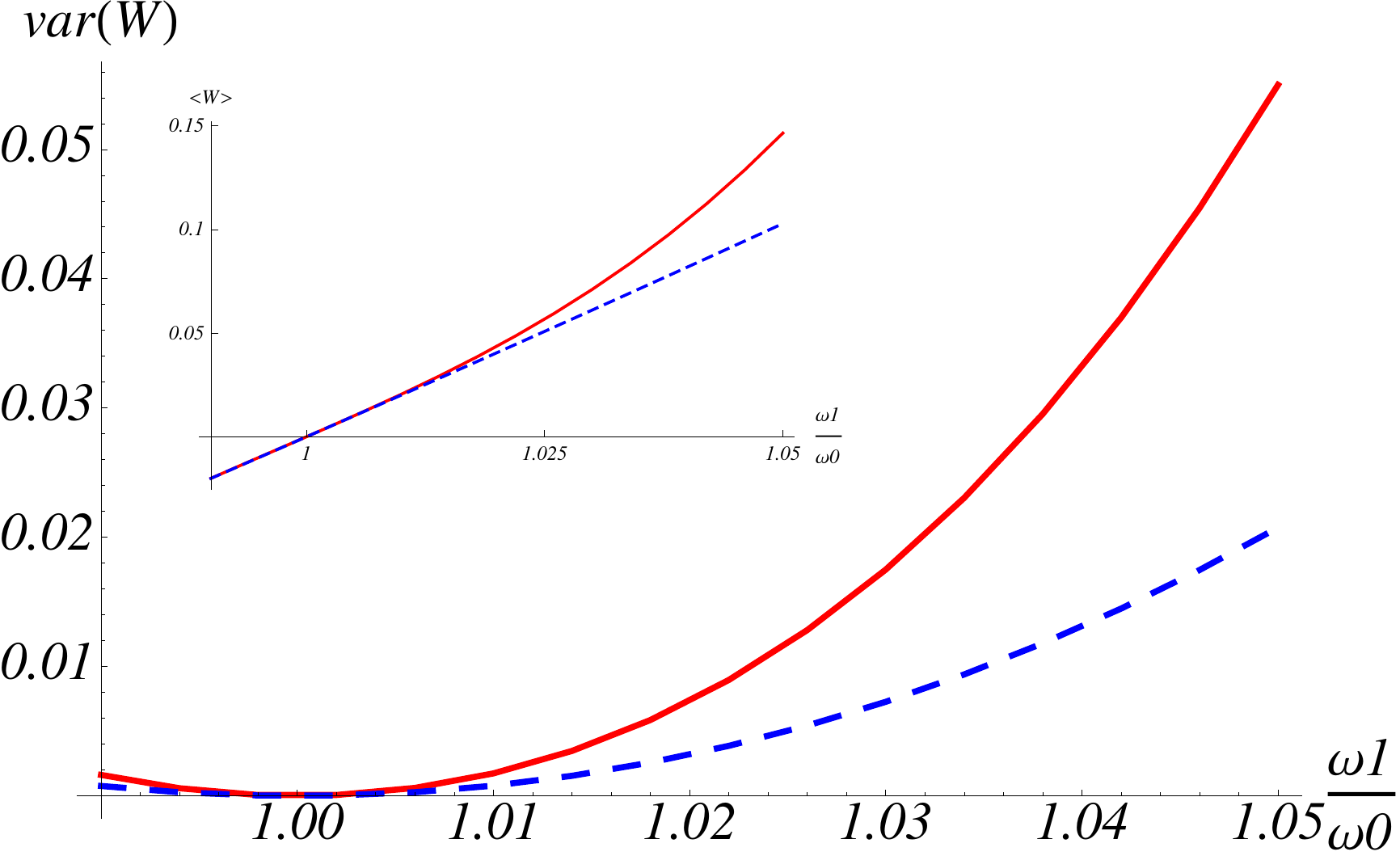}
\caption{Variance and mean (inset) work for an oscillator with weak anharmonic corrections (red) (\ref{eq41})  compared with those of the unperturbed oscillator (blue, dashed) (\ref{eq2}) ($\omega_0=0.5$, $\tau=1$, $\beta=0.5$, $M=1$, $\hbar=1$ and $\sigma_\alpha=0.025$). \label{fig5}}
\end{figure}
 For the numerical analysis, we choose the parameterization of $\omega^2_t$ to be linear in time,
\begin{equation}
\label{eq44}
\omega_t^2=\omega_0^2+\left(\omega^2_1-\omega_0^2\right)\frac{t}{\tau}\,.
\end{equation}
Since the anharmonic corrections are given by the geometric set-up and hence directly scale with the angular frequency of the harmonic oscillator, we assume that 
\begin{equation}
\label{eq45}
\alpha_t= \sigma_\alpha\left(\omega^2_1-\omega_0^2\right)\frac{t}{\tau}\,.
\end{equation}
The parameter $\sigma_\alpha$ controls the strength of the perturbation. 
In Fig. \ref{fig5}, we have plotted the mean work $\la W\ra$ and the variance, $var(W)=\la W^2\ra-\la W\ra^2$, of the work distribution for the anharmonically perturbed harmonic oscillator (\ref{eq41}), together with the exact result for the unperturbed oscillator (\ref{eq2}).  We observe that both quantities are enhanced by the anharmonicity. From the analytical expressions of the transition probabilities (\ref{C3}), we see that additional transitions, $m=n\pm4$, now become possible because of the quartic correction. These additional transitions lead to a larger mean and variance of the work. Based on our discussion in section \ref{measure}, we can therefore conclude that the anharmonic perturbation increases the degree of nonadiabaticity of the frequency change. Numerical comparison further shows that the effect of $A_t$ can be neglected up to a strength of roughly one percent, $\sigma_\alpha\lesssim0.01$, of the harmonic amplitude $\Omega_t$. For a standard trap configuration with trap frequencies of the order kHz-MHz, the harmonic assumptions is fulfilled up to energies of the order of eV, see Ref.~\cite{hub08a}, and the effect of anharmonic corrections are negligible for these energies.

\subsection{Random electric field corrections \label{thermal}}
Linear Paul traps are almost perfectly isolated from their surroundings. They however suffer from the presence of random electric fields that are generated in the trap electrodes \cite{tur00}. These weak fluctuating fields are the source of motional heating of the charged ions confined in the harmonic trap.
The  Hamiltonian of the quantum oscillator in the presence of the field is 
\begin{equation}
\label{eq46}
H_t = H_0 + \Omega_t + \Lambda_t
\end{equation}
where  the small perturbation $\Lambda_t$ is linear in position,
\begin{equation}
\label{eq47}
\Lambda_t =\lambda_t x\,.
\end{equation}
The function  $\lambda_t= q E_t$ is proportional to the random electric field $E_t$ ($q$ is the charge of the ion) and is taken to be Gaussian distributed  with
\begin{equation}
\label{eq48}
\la \lambda_t\ra=0\hspace{2em}\text{and}\hspace{2em}\la\lambda_t\,\lambda_s\ra= \kappa_{t,s}\,.
\end{equation}
The heating rate of the trap is related to the spectral density of the noise $\lambda_t$ \cite{sav97}
\begin{equation}
\label{eq49}
\la \dot n\ra\simeq\frac{1}{4\,M\,\hbar \omega_t}\,\int\limits_{-\infty}^{+\infty} ds \,\exp{\left(i \omega_t s\right)\,\la \lambda_t\,\lambda_{t+s}\ra}\,.
\end{equation}
We first  calculate the transition probabilities $p_{m,n}^{\tau}$ for a fixed value of  $\lambda_t$ and then average over $\lambda_t$ using Eq.~\eqref{eq48}. In complete analogy  to Eq.~(\ref{eq43}), we obtain
\begin{equation}
\label{eq50}
p_{m,n}^{\tau} = \left|\delta_{m,n} + \frac{1}{i\hbar} \int\limits_0^{\tau} dt \, \exp{\left(i \omega_{m,n} \, t\right)} \, (\Omega_{m,n}^t + \Lambda_{m,n}^t) \right|^2\,,
\end{equation}
with the interaction matrix elements $\Lambda_{m,n}^t$ given by,
\begin{equation}
\label{eq51}
\Lambda_{m,n}^t = \lambda_t \sqrt{\frac{\hbar}{2\,M \, \omega_0}} \, \left(\sqrt{n+1} \, \delta_{m,n+1} + \sqrt{n} \, \delta_{m,n-1} \right)\,.
\end{equation}
The explicit expression of the transition probabilities can be found in \ref{thermal_pert}.
After averaging over all possible $\lambda_t$, the transition probabilities can be divided into two distinct   contributions coming from the parametric variation of the frequency ($\Omega_t$ in Eq.~\eqref{eq46}) and  the noise term ($\Lambda_t$ in Eq.~\eqref{eq46}),
\begin{equation}
\label{eq52}
\la
p_{m,n}^\tau\ra_{\lambda_t}=p_{m,n}^\tau(\omega_t)+p_{m,n}^\tau\left(\la\lambda_t\,\lambda_s\ra\right)\,.
\end{equation}
Similarly, we can separate the mean final energy into a deterministic and a
stochastic  part,
\begin{equation}
\label{eq53}
\begin{split}
\la H_\tau\ra&=\sum\limits_{m,n} E_m^\tau\,
\left(p_{m,n}^\tau(\omega_t)+p_{m,n}^\tau\left(\la\lambda_t\,\lambda_s\ra\right)
\right)\, p_n^0\\
&=\frac{\hbar\omega_1}{2}\, \left(Q^*+Q^*_{\lambda_t}\right)
\coth{\left(\frac{\beta}{2}\,\hbar\omega_0\right)}\,.
\end{split}
\end{equation}
Here the parameter $Q^*_{\lambda_t}$ is defined  as
\begin{equation}
Q^*_{\lambda_t}=\frac{\la H_\tau\ra_{\lambda_t}}{\hbar\omega_1/2\,\coth{\left(\beta/2\,\hbar\omega_0\right)}}
\end{equation}
with
\begin{equation}
\la H\ra_{\lambda_t}=\sum\limits_{n,m} \hbar\omega_1 \left(m+\frac{1}{2}\right)\,p_{m,n}^\tau\left(\la\lambda_t\,\lambda_s\ra\right)\,p_n^0\,.
\end{equation}
\begin{figure}
\centering
\includegraphics[width=0.47\textwidth]{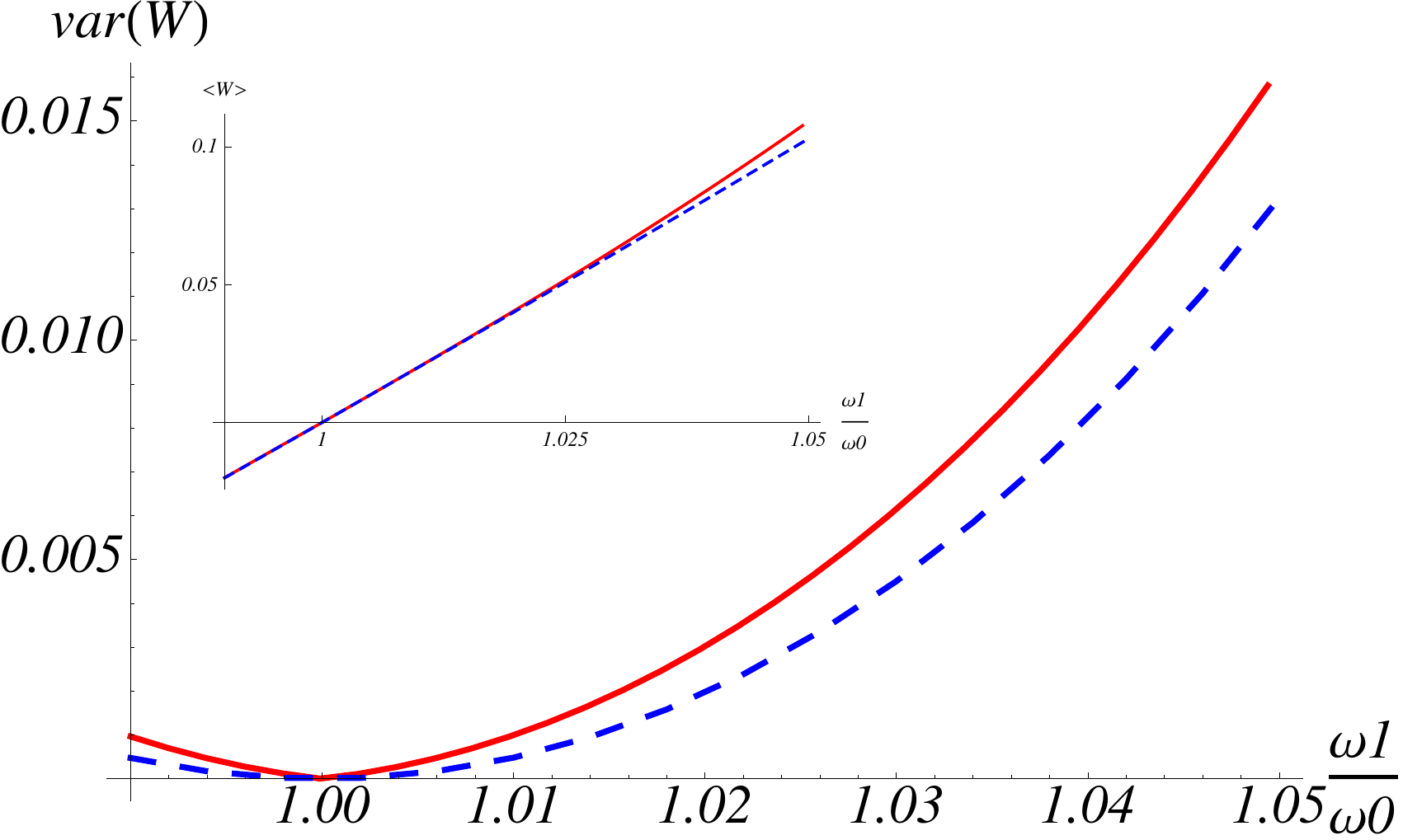}
\caption{Variance and mean (inset) work for a charged oscillator with weak electric noise (red) (\ref{eq41})  compared with those of the unperturbed oscillator (blue, dashed) (\ref{eq2}) ($\omega_0=0.5$, $\tau=1$, $\beta=0.5$, $M=1$, $\hbar=1$ and $\sigma_\alpha=0.025$).\label{fig6}}
\end{figure}
Equation \eqref{eq43} shows that the effect of the random electric field is to renormalize the adiabaticity parameter $Q^* \rightarrow Q^*+Q^*_{\lambda_t}$. Both the mean and the variance of the work distribution  are increased as depicted in Fig.~\ref{fig6}.
The fluctuating field thus enhances the degree of nonadiabaticity. This effect can be understood by noting that the perturbation $\Lambda_t$ generates additional transitions between states (the latter obey $m=n\pm1$). 
We observe that the variance is more sensitive to the perturbation than the mean, since it depends quadratically on $Q^*$ and not linearly. For the numerical analysis
we have chosen a  white noise of the form,
\begin{equation}
\label{eq54}
\kappa_{t,s}=\sigma_\lambda\,\left(\omega_0^2-\omega_1^2\right)\,\delta(t-s)\,,
\end{equation}
where the relative noise strength is given by $\sigma_\lambda$. As for the case of the anharmonic perturbation, we note that  one can neglect the influence of the electric noise up to a relative strength of roughly one percent,
$\sigma_\lambda\lesssim0.01$. 

\section{Conclusion} 
We have considered the statistics of the nonequilibrium work of a quantum oscillator when its angular frequency is varied in time. Due to the quantized nature of the energy spectrum, the work probability distribution is discrete. We have analyzed the discrete-to-continuous transition of the work distribution in various limits by introducing the cumulative function. We have shown that the cumulative work distribution becomes smooth in the limit of high temperatures and of large values of the adiabaticity parameter $Q^*$; in both regimes, the mean energy of the oscillator is much larger than the energy separation, and the spectrum can be considered quasi-continuous. We have moreover developed a perturbative approach to investigate the effects of small quartic anharmonicities on the work distribution. We  have found that the latter increase both the mean and the variance of the final energy of the oscillator, indicating an augmentation of the nonadiabaticity of the frequency change. In a similar way, we have studied the influence of a weak electric noise on a charged harmonic oscillator and obtained an analogous enhancement of the degree of nonadiabaticity. Our results permit an accurate description of measured quantum work distributions in modulated Paul trap under realistic experimental conditions.

\subparagraph{Acknowledgements}
We would like to thank Peter  Talkner for discussions and comments. This work is based on OA's project within the NIM summer research program, and was supported by the Emmy Noether Program of the DFG (contract LU1382/1-1) and the cluster of excellence Nanosystems Initiative Munich (NIM).

\appendix

\section{Analytical expression for $Q^*$\label{Q*}}

Closed expressions for the adiabaticity parameter $Q^*$ can be found whenever the classical  equation (\ref{eq7})
\begin{equation}
\label{A1}
\frac{d^2}{dt^2}\,X_t+\omega^2_t\,X_t=0 
\end{equation}
can be solved analytically. Equation (\ref{A1}) is of the general form  of a Hill equation which can be solved under various conditions \cite{tak86a,tak86}. Equation (\ref{A1}) reduces to common differential equations in the case of specific parameterizations  $\omega_t$. Thus  for the case of linear parameterizations (\ref{eq44}), the solutions are given in terms of the Airy-functions \cite{def08}. On the other hand, for a sinusoidal parameterization
\begin{equation}
\label{A2}
\omega_t^2=\omega_1^2-\left(\omega_1^2-\omega_0^2\right)\, \cos{ \left(\frac{\pi}{2}\,\frac{t}{\tau}\right)}
\end{equation}
Eq.~(\ref{A1}) takes the form of the Mathieu equation. The solutions $X_t$ and $Y_t$ can then be written as,
\begin{equation}
\label{A3}
\begin{split}
X_t=&\frac{4 \tau}{\pi}\,  \Bigg(\text{S}\left[d_1,d_2,0\right]\text{C}\left[d_1,d_2,\frac{\pi  t}{4 \tau }\right]\\
-&\text{C}\left[d_1,d_2,0\right]\text{S}\left[d_1,d_2,\frac{\pi  t}{4 \tau }\right]\Bigg)\\
\times& \Bigg(\text{C'}\left[d_1,d_2,0\right]\text{S}\left[d_1,d_2,0\right]\\
-&\text{C}\left[d_1,d_2,0\right]\text{S'}\left[d_1,d_2,0\right]\Bigg)^{-1}
\end{split}
\end{equation}
and
\begin{equation}
\label{A4}
\begin{split}
Y_t=& \Bigg(\text{S'}\left[d_1,d_2,0\right]\text{C}\left[d_1,d_2,\frac{\pi  t}{4 \tau}\right]\\
-&\text{C'}\left[d_1,d_2,0\right] \text{S}\left[d_1,d_2,\frac{\pi  t}{4 \tau}\right]\Bigg)\\
\times&\Bigg(\text{C}\left[d_1,d_2,0\right]\text{S'}\left[d_1,d_2,0\right]\\
-&\text{C'}\left[d_1,d_2,0\right]\text{S}\left[d_1,d_2,0\right]\Bigg)^{-1}\,.
\end{split}
\end{equation}
Here the functions C and S  denote the corresponding Mathieu functions \cite{abra}. The parameters $d_1$ and $d_2$ are given by,
\begin{equation}
\label{A5}
d_1=\frac{16}{\pi^2}\,\omega_1^2\tau^2
\end{equation}
and
\begin{equation}
\label{A6}
d_2=\frac{8}{\pi^2}\,\left(\omega_1^2-\omega_0^2\right)\,\tau^2\,.
\end{equation}
Further analysis of the parameter $Q^*$ in the context of vacuum squeezing can be found in Ref.~\cite{gal09}.

\section{Exact transition probabilities\label{trans}}

We here collect the analytical expressions of the transition probabilities $p_{m,n}^\tau$ \cite{hus53}. Despite its apparent simplicity, the generation function $P(u,v)$ (\ref{eq13}) cannot be expanded in powers of $u$ and $v$ in an exact series. We thus make use of the $p_{m,n}^\tau$ as defined by the matrix elements of the propagator $U_{\tau,0}(x|x_0)$, $p_{m,n}^{\tau}=\left|U_{m,n}^\tau\right|^2$ (\ref{eq10}). This matrix elements are given by
\begin{equation}
\label{B1}
 U_{m,n}^\tau=\int dx_0\int dx \, {\phi^*}_m^\tau(x) U_{\tau,0}(x|x_0)\phi^0_n(x_0).
\end{equation}
We use again the method of generating functions. We use the linear generating function of $\phi^t_n(x)$ \cite{coh1},
\begin{equation}
\label{B2}
\begin{split}
 \sum\limits_{n=1}^{\infty}\left(\frac{\sqrt{\pi}\,2^n}{n!}\right)^{1/2}\,z^n {\phi_n^t}(x)=\hspace{2cm}\\
\sqrt[4]{\frac{M \omega_t}{\hbar}}\exp{\left(-{\frac{M\omega_t}{2\hbar}}\,x^2+2\sqrt{\frac{M \omega_t}{\hbar}}\,z\,x-z^2\right)} \ ,
\end{split}
\end{equation}
to evaluate the generating function of the propagator, 
\begin{equation}
\label{B3}
 U(u,v)=\sum\limits_{m,n} \left(\frac{\pi\,2^{n+m}}{n!\,m!}\right)^{1/2}\,u^n v^m U_{m,n}^\tau \ .
\end{equation}
By introducing the complex parameters,
\begin{eqnarray}
\label{B4}
 \zeta  &=&  \omega_1\omega_0 X_\tau -\omega_0 i \dot X_\tau+\omega_1 i Y_\tau +\dot Y_\tau\nonumber\\
 |\zeta|^2 &=&  2\omega_0\omega_1\,(Q^*-1) \\
\label{B5}
 \sigma &=& \omega_1\omega_0 X_\tau-\omega_0 i \dot X_\tau -\omega_1 i Y_\tau -\dot Y_\tau\nonumber\\
|\sigma|^2  &=&   2\omega_0\omega_1(Q^*+1)
\end{eqnarray}
we can write
\begin{equation}
\label{B6}
 U(u,v)=\frac{\sqrt[4]{\omega_0\omega_1}}{\sqrt{i\sigma/2\pi}}\exp{\left(\frac{\zeta u^2-4i\sqrt{\omega_0\omega_1}uv+{\zeta}^* v^2}{\sigma}\right)}.
\end{equation}
The matrix elements $U_{m,n}^\tau$ can then be obtained by a series expansion of (\ref{B6}) in powers of $u$ and $v$ \cite{hus53}
\begin{equation}
\label{B7}
\begin{split}
U_{m,n}^\tau=&\sqrt[4]{2 \omega_0\omega_1} \sqrt{\frac{n!\,m!\,\zeta^n {\zeta^*}^m }{2^{n+m-1}i\sigma^{n+m+1}}}\\
\times&\sum\limits_{l=0}^{\min{(m,n)}}\frac{[-2 i \sqrt{2/(Q^*-1)}]^l}{l!\,[(n-l)/2]!\,[(m-l)/2]!} \ .
\end{split}
\end{equation}
According to the selection~rule $m=n\pm2k$, $l$ runs over even numbers only, if $m$, $n$ are even, and over odd numbers only, if $m$, $n$ are odd. The explicit expression for the matrix elements  $U_{m,n}^\tau$ then reads for even elements
\begin{equation}
\label{B8}
\begin{split}
 U_{2\mu,2\nu}^\tau=&\sqrt{\frac{2\nu!2\mu!}{2^{2\nu+2\mu-1}i}}\\
\times&\sqrt{\frac{\zeta^{2\nu} {\zeta^*}^{2\mu}}{\sigma^{2\nu+2\mu+1}}}\frac{\sqrt[4]{2 \omega_0\omega_1}}{\Gamma(\mu+1)\,\Gamma(\nu+1)}\\
\times&_2 F_1\left(-\mu,\,-\nu;\,\frac{1}{2};\,\frac{2}{1-Q^*}\right)
\end{split}
\end{equation}
and for odd elements
\begin{equation}
\label{B9}
\begin{split}
 U_{2\mu+1,2\nu+1}^\tau=&-\sqrt{\frac{8i\,(2\nu+1)!(2\mu+1)!}{(Q^*-1)\,2^{2\nu+2\mu+1}}}\\
\times&\sqrt{\frac{\zeta^{2\nu+1}{\zeta^*}^{2\mu+1}}{\sigma^{2\nu+2\mu+1}}}\frac{\sqrt[4]{2 \omega_0\omega_1}}{\Gamma(\mu+1)\,\Gamma(\nu+1)}\\
\times&_2 F_1\left(-\mu,\,-\nu;\,\frac{3}{2};\,\frac{2}{1-Q^*}\right).
\end{split}
\end{equation}
We have here introduced the hypergeometric function $_2 F_1$ \cite{abra} in order to simplify the sums and write the matrix elements $U_{m,n}^\tau$ in closed form. $\Gamma(x)$ denotes the Euler Gamma function. Combining everything, we get the explicit expressions for the transition probabilities which reads for even transitions
\begin{equation}
\label{B10}
\begin{split}
p_{2\mu,2\nu}^{\tau}=&\frac{2^{1/2}}{(Q^*+1)^{1/2}}\,\left(\frac{Q^*-1}{Q^*+1}\right)^{\mu+\nu}\\
\times&\frac{\Gamma(1/2 +\mu)\,\Gamma(1/2+\nu)}{\pi\Gamma(1+\mu)\,\Gamma(1+\nu)}\\
\times&\left[_2F_1\left(-\mu,\,-\nu;\,\frac{1}{2};\,\frac{2}{1-Q^*}\right)\right]^2
\end{split}
\end{equation}
and for odd transitions
\begin{equation}
\label{B11}
\begin{split}
p_{2\mu+1,2\nu+1}^{\tau}=&\frac{2^{7/2}}{(Q^*+1)^{3/2}}\,\left(\frac{Q^*-1}{Q^*+1}\right)^{\mu+\nu}\\
\times&\frac{\Gamma(3/2 +\mu)\,\Gamma(3/2+\nu)}{\pi\Gamma(1+\mu)\,\Gamma(1+\nu)}\\
\times&\left[_2F_1\left(-\mu,\,-\nu;\,\frac{3}{2};\,\frac{2}{1-Q^*}\right)\right]^2. 
\end{split}
\end{equation}

\section{Perturbational transition probabilities\label{trans_per}}

In this appendix, we provide  the approximate first-order transition probabilities calculated with the help of time-dependent perturbation theory.

\subsection{Isolated harmonic oscillator}

Following Eqs.~(\ref{eq39}) and (\ref{eq40}), the transition probabilities for an isolated harmonic oscillator are
\begin{equation}
\label{C1}
\begin{split}
 p_{m,n}^\tau=&\Bigg| \delta_{m,n} + \frac{i}{4\omega_0} \int\limits_0^{\tau} dt \, \exp{\left(i\omega_{m,n}  t\right)}\left(\omega_0^2 - \omega^2_t\right)\\
\times&\bigg[ \sqrt{n + 1}\sqrt{n+2} \, \delta_{m,n+2} + (2n +1) \, \delta_{m,n}\\ 
+&\sqrt{n}\sqrt{n-1} \, \delta_{m,n-2}\bigg]\Bigg|^2\,.
\end{split}
\end{equation}
The  selection rule $m=n\pm2$ finds its origin in the presence of the Kronecker deltas.

\subsection{Anharmonic corrections}

The interaction matrix element $A_{m,n}^t$ (\ref{eq42}) are given by
\begin{equation}
\label{C2}
\begin{split}
A_{m,n}^t =& \alpha_t \left(\frac{\hbar}{2\, M \omega_0}\right)^2 \\
\times&\bigg[ \sqrt{n+1} \sqrt{n+2} \sqrt{n+3} \sqrt{n+4} \, \delta_{m,n+4}\\
+&   \left(4 n + 6\right) \sqrt{n+1} \sqrt{n+2} \, \delta_{m,n+2}\\
+&  \left(6 n^2+6 n +3\right) \, \delta_{m,n} \\
+& \left(4n-2\right) \sqrt{n} \sqrt{n-1} \, \delta_{m,n-2}\\
+&  \sqrt{n} \sqrt{n-1} \sqrt{n-2} \sqrt{n-3} \, \delta_{m,n-4} \bigg]\,.
\end{split}
\end{equation}
The transition probabilities (\ref{eq43}) for a harmonic oscillator with quartic  corrections then read
\begin{equation}
\label{C3}
\begin{split}
p_{m,n}^{\tau} =& \Bigg| \delta_{m,n} + \frac{1}{i\hbar} \int\limits_{0}^{\tau}  dt \, \exp{\left(i \omega_{m,n} \, t\right)}\\ 
\times& \Bigg\{ -\frac{\hbar}{4 \omega_0}\, \left(\omega_0^2 - \omega^2_t\right) \bigg[\sqrt{n + 1}\sqrt{n+2} \, \delta_{m,n+2} \\
+& \, (2n +1) \, \delta_{m,n} +  \sqrt{n}\sqrt{n-1} \, \delta_{m,n-2}\bigg]\\
+& \alpha_t \left(\frac{\hbar}{2 M \omega_0}\right)^2\\
\times& \bigg[ \sqrt{n+1} \sqrt{n+2} \sqrt{n+3} \sqrt{n+4} \, \delta_{m,n+4} \\ 
+&  \left(4 n + 6\right) \sqrt{n+1} \sqrt{n+2} \, \delta_{m,n+2} \\
+&  \left(6 n^2+6 n +3\right)\,\delta_{m,n} \\
+&  \left(4n-2\right) \sqrt{n} \sqrt{n-1}  \, \delta_{m,n-2} \\ 
+&  \sqrt{n} \sqrt{n-1} \sqrt{n-2} \sqrt{n-3} \,\delta_{m,n-4} \bigg] \Bigg\}\Bigg|^2 
\end{split}
\end{equation}
where additional transitions $m=n\pm4$ become possible.

\subsection{Random electric field corrections\label{thermal_pert}}

In the presence of an external electric field (\ref{eq47}), the transition probabilities (\ref{eq50}) become
\begin{equation}
\label{C4}
\begin{split}
p_{m,n}^{\tau} =& 
\Bigg| \delta_{m,n} + \frac{1}{i\hbar} \int\limits_0^{\tau} dt \, \exp{\left(i\omega_{m,n}  t\right)}\\ 
\times&\Bigg\{ - \frac{ \hbar}{4 \omega_0}\,\left(\omega_0^2 - \omega^2_t\right)\,\bigg[ \sqrt{n + 1}\sqrt{n+2} \, \delta_{m,n+2}\\
+& \left(2n +1\right) \, \delta_{m,n} + \sqrt{n}\sqrt{n-1} \, \delta_{m,n-2}\bigg]\\
+& \lambda_t  \sqrt{\frac{\hbar}{2 M\omega_0}} \, \bigg(\sqrt{n+1} \, \delta_{m,n+1} + \sqrt{n} \, \delta_{m,n-1} \bigg)\Bigg\}\Bigg|^2\,.
\end{split}
\end{equation}
The Kronecker deltas now also allow next-neighbor transitions $m=n\pm1$. Averaging over the noise $\lambda_t$ using Eq.~(\ref{eq48}), we finally obtain,
\begin{equation}
\label{C5}
\begin{split}
\la p_{m,n}^{\tau}\ra_{\lambda_t} =& 
 \Bigg| \delta_{m,n} + \frac{1}{i\hbar} \int\limits_0^{\tau} dt \, \exp{\left(i\omega_{m,n}  t\right)}\\ 
\times&\Bigg\{ - \frac{ \hbar}{4 \omega_0}\,\left(\omega_0^2 - \omega^2_t\right)\,\bigg[ \sqrt{n + 1}\sqrt{n+2} \, \delta_{m,n+2}\\
+& \left(2n +1\right) \, \delta_{m,n} + \sqrt{n}\sqrt{n-1} \, \delta_{m,n-2}\bigg]\Bigg\}\Bigg|^2\\
+&\frac{1}{2 M\,\hbar\omega_0}\,\left[\left(n+1\right) \, \delta_{m,n+1} + n \, \delta_{m,n-1} \right]\\
\times&\left| \int\limits_0^{\tau} dt\int\limits_0^{\tau} ds \, \exp{\left(i\omega_{m,n}  \left(t-s\right)\right)}\,\la\lambda_t\,\lambda_s\ra \right|\,.
\end{split}
\end{equation}

\end{document}